\documentclass[aps,prb,twocolumn,groupedaddress,floatfix]{revtex4-1}

\usepackage{graphicx}
\usepackage{dcolumn}
\usepackage{bm}

\begin{document}

\title{Spin relaxation in  strained  graphene nanoribbons: armchair vs zigzag edges}
\author{Sanjay Prabhakar,  Roderick Melnik and Shyam Badu}
\affiliation{
The MS2Discovery Interdisciplinary Research Institute, M\,$^2$NeT Laboratory, Wilfrid Laurier University, Waterloo, ON, N2L 3C5 Canada
}
\date{August 21, 2014}

\begin{abstract}
 We study the influence of ripple waves originating from the electromechanical effects  on  spin relaxation caused by electromagnetic fields in armchair and zigzag graphene nanoribbons (GNRs). By utilizing  analytical expressions supported by numerical simulations, we show that it is possible to tune the spin flip behaviors ON and OFF due to ripple waves in GNRs for potential applications in straintronic devices. This finding is  similar to recently made observations on the design of spintronic devices in III-V semiconductor quantum dots, where the sign change in the effective Land$\mathrm{\acute{e}}$ $g$-factor can be engineered with the application of gate controlled electric fields.  In particular,  we show that  the tuning of spin extends to larger widths for the armchair GNRs than for the zigzag GNRs. Here we also report that   the relaxation rate vanishes like $L^5$.
\end{abstract}



\maketitle

\section{Introduction}

Graphene has attracted potential interest for   the design of optoelectronic devices because it possesses  unique  electronic and physical properties  due to the presence of Dirac-like energy spectrum of the charge
carriers.~\cite{dassarma11,nato09,prabhakar13,droth13,droth11,novoselov05a,abanin06} Several observed quantum phenomena such as the half integer quantum Hall effect, non-zero Berry phase, as well as the  measurement of conductivity of electrons in the electronic devices lead to novel applications in carbon based  nanoelectronic devices.~\cite{abanin06,dassarma11,nato09,droth13,droth11}
The experimental data observed by the quantum Hall effect measurement technique  suggests a one atom thick graphene sheet has the same properties as a two dimensional system that does not contain any bandgap at two Dirac points.~\cite{novoselov05a,abanin06}
In addition, the  researchers around the globe   desire  to build next generation semiconductor devices from graphene  because the experimental data shows that this material possess the high mobility of charge carriers. In such devices, one finds  an opportunity to control electronic properties of graphene-based structures
using several different techniques such as gate controlled electric fields and  magnetic fields. Further one can engineer the straintronic devices by controlling the electromechanical properties  via the  pseudomorphic gauge fields.~\cite{droth13,droth11,nato09,shenoy08,choi10,maksym13,bao12,balandin08,
yan12,shahil12,cadelano09,cadelano12,bao-lau09,shahil12}

Two dimensional images of graphene sheet taken from high resolution transmission electron microscope or scanning tunneling microscope show that its  surface normal  varies by several degrees and the out-of-plane deformations reach to the nanometer scale that is considered to be due to the presence of  ripple waves in graphene sheet.~\cite{meyer07,bonilla12,guinea10}
Ripples in graphene  are induced by several different mechanisms that  have been widely investigated.~\cite{gibertini10,shenoy08,wang12,barbier10,ferreira11,lim12,cadelano09,choi10,gui08,kitt12,ramasubramaniam12,
shenoy08,wang12,bonilla12,carpio08,zhou08,sanjose11,sanjose12,cadelano12,balandin08,shahil12,gibertini10,duhee11,lajevardipour12}
Such ripples  are part of the  intrinsic properties of graphene that are expected to  strongly affect the  bandstructures  due to their  coupling through pshedomorphic vector potential.~\cite{bao-lau09,bao12,cerda-mahadevan03}
Monte-Carlo simulations results   show that the  ripples spontaneously appear owing to thermal fluctuations with a size distribution due to the multiplicity of chemical bonding in carbon.~\cite{fasolino07} Recent experimental studies on graphene  at several different annealed temperatures  confirmed that the amplitude of the  ripple waves is enhanced with increasing temperature.~\cite{bao-lau09}
At the same time, recent theoretical studies show that ripples might be induced due to spontaneous symmetry breaking by flexural phonons.~\cite{sanjose11} The phonon induced ripples  follow a mechanism similar to that responsible for the condensation of the Higgs field in relativistic field theory leading to a zero-temperature buckling transition in which the order parameter is given by the square of the gradient of the flexural phonon field. In experiments on graphene  suspended on substrate trenches, there appear much longer and taller waves (close to a micron scale) directed parallel to the applied stress. These long wrinkles can also be induced thermally.~\cite{bao-lau09}  In this paper we present a model that couples the Navier equations, accounting for electromechanical effects,  to the electronic properties of armchair and zigzag graphene nanoribbons. We show that the ripple waves originating from the electromechanical effects strongly influence the bandstructure of GNRs. This response mechanism might be used for tuning the spin currents ON and OFF  in the spin relaxation behaviors caused by electromagnetic fields for potential applications in straintronic devices.

The paper is organized as follows: In section~\ref{theoretical-model}, we provide a theoretical description of coupling between the electromechanical effects and the bandstructure  of armchair and zigzag GNRs. Here we also provide an analytical expression of the energy spectrum of GNRs with the armchair and zigzag edges. Numerical schemes for obtaining electromechanical effects on the bandstructure calculation of GNRs via  exact diagonalization technique is discussed in section~\ref{computational-details}. In addition to the results associated to the electromechanical and bandstructure of GNRs in section~\ref{results-discussions}, we also estimate the electromagnetic field mediated spin relaxation time in GNRs. Finally we summarize our results in section~\ref{conclusion}.

\begin{figure}
\includegraphics[width=8cm,height=8cm]{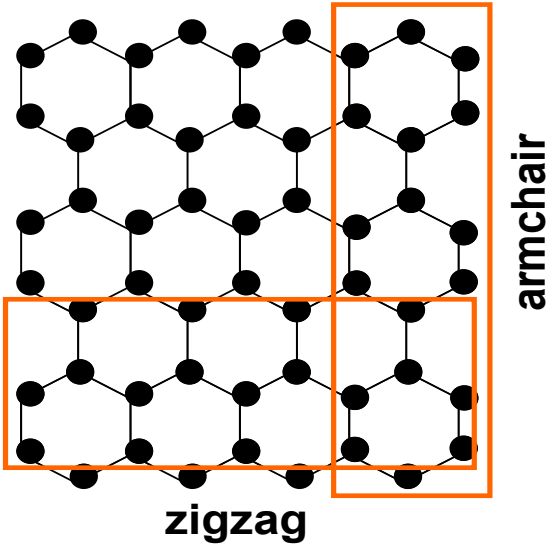}
\caption{\label{fig1}
The lattice structure of hexagon graphene sheet with the armchair and zigzag edges. The  vertical and horizontal recangle correspond to the zigzag and armchair graphene nanoribbons.
}
\end{figure}
\begin{figure*}
\includegraphics[width=16cm,height=10cm]{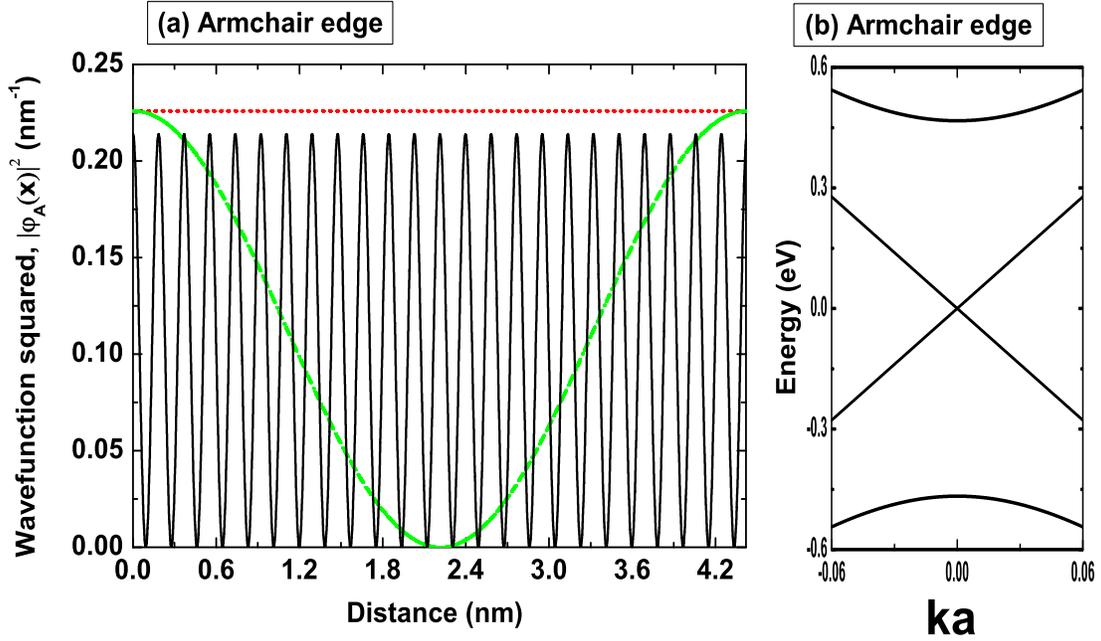}
\caption{\label{fig2}
(a) Wavefunctions squared vs distance ($x$) in GNR of armchair edge  at Dirac $K$ point ($k_y=0$).  We choose the width of the GNR as $L=3\sqrt{3} ~a N$ ($N=6$) (dotted line for $n=4N$ and dashed-dotted line for $n=4N+1$) and $L=\sqrt{3} ~a \left(3N+1\right)$ (solid line for $n=0$) to reproduce the results of Ref.~(\onlinecite{brey06}).  (b) The bandstructures of metallic (i.e. $L=3\sqrt{3} ~a N$ ($N=6$)) GNR with the armchair edge. We also choose $v_F=10^6~m/s$.
}
\end{figure*}
\begin{figure*}
\includegraphics[width=16cm,height=8cm]{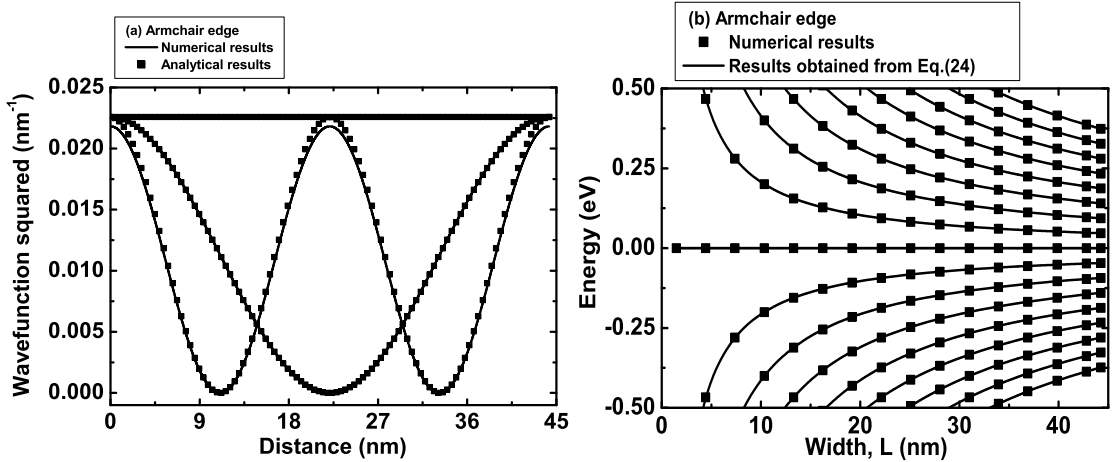}
\caption{\label{fig3}
(a) Ground and first excited state wavefunctions squared vs distance ($x$) in GNR of armchair edge  at Dirac $K$ point ($k_y=0$). Here we consider $L=3\sqrt{3} ~a N$ ($N=60$). (b) Several energy eigenvalues vs width of the GNR of armchair edge at Dirac $K$ points ($k_y=0$). For analytical results, we first add Eqs.~(\ref{phi-A-arm}) and~(\ref{phi-p-A}) (i.e., $\tilde{\phi}_A=\phi_A+\phi'_A$) to admix the valley and then found the wavefunctions squared from the expression: $|\tilde{\phi}_A|^2=\cos^2\left(k_nx\right)/L$.  Note that the numerical results obtained via exact diagonalization technique are in excellent agreement with the analytical results.
}
\end{figure*}
\begin{figure*}
\includegraphics[width=16cm,height=10cm]{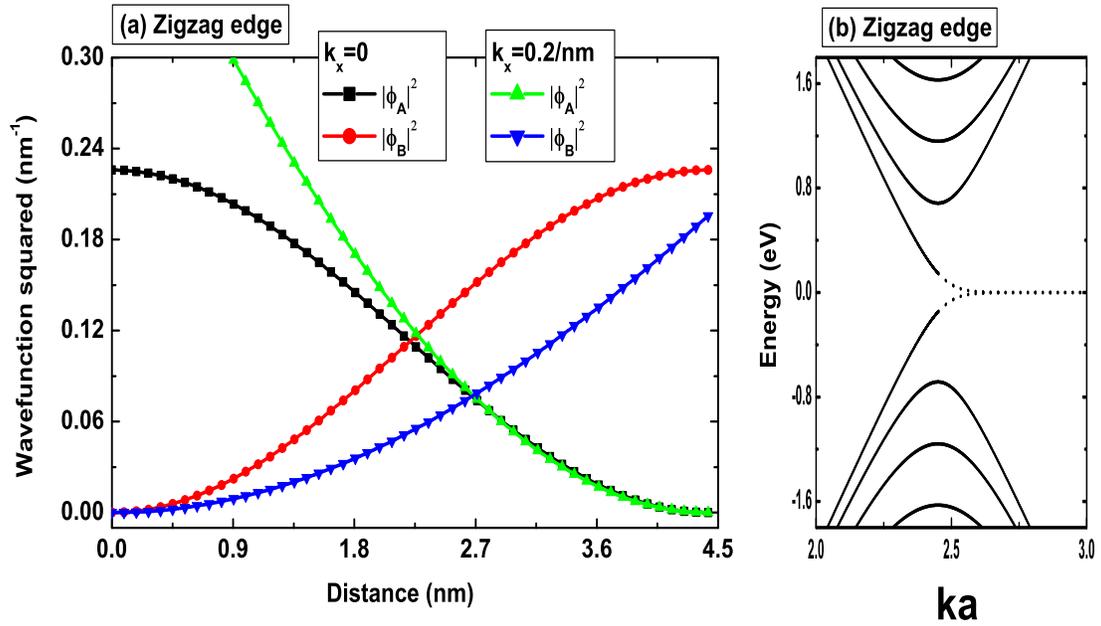}
\caption{\label{fig4}
(a) Ground state wavefunctions squared vs distance ($y$) in GNR of zigzag edge.  These plots are obtained from Eqs.~(\ref{phi-A}) and~(\ref{phi-B}). (b) The bandstructures of  GNR with the zigzag edge are obtained from Eqs.~(\ref{varepsilon-2}) (dotted lines) and~(\ref{varepsilon-3}) (solid lines).
Here we choose the width of the GNR as $L=3\sqrt{3} ~a N$ ($N=6$) to reproduce the results of Ref.~\onlinecite{brey06}.
}
\end{figure*}
\begin{figure*}
\includegraphics[width=16cm,height=8cm]{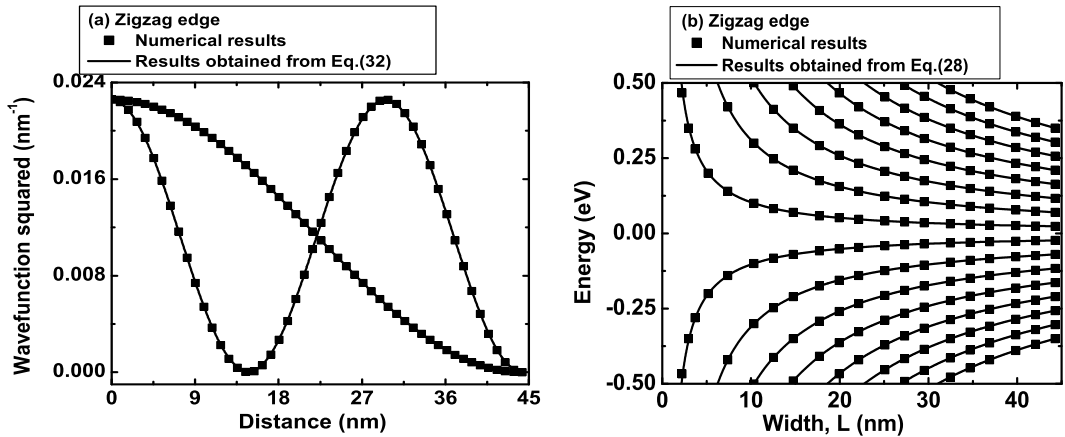}
\caption{\label{fig5}
(a) Ground and first excited state wavefunctions squared vs distance ($y$) in zigzag GNRs at Dirac $K$ point ($k_x=0$). Here we consider $L=3\sqrt{3} ~a N$ ($N=60$). (b) Several energy eigenvalues of electron and hole like states vs width of the zigzag GNRs at Dirac $K$ points ($k_x=0$). Note that the numerical results obtained via exact diagonalization technique are in excellent agreement with the analytical results.
}
\end{figure*}
\begin{figure}
\includegraphics[width=8.5cm,height=8cm]{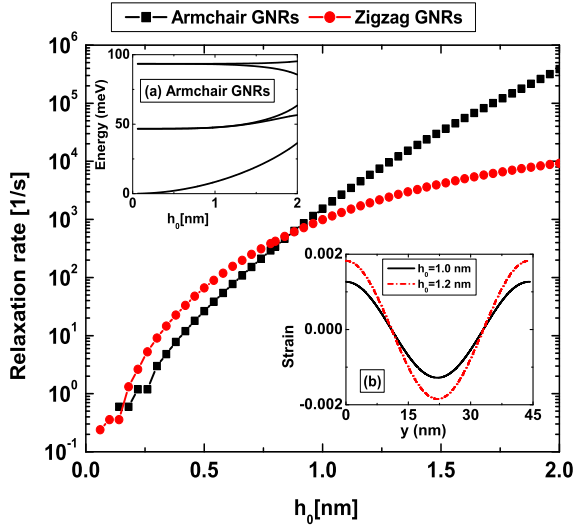}
\caption{\label{fig6}
Electromagnetic field mediated spin relaxation vs height of the ripple waves at Dirac point in armchair  and zigzag GNRs. Evidently in inset plot of Fig.~\ref{fig6}(a), at $h_0 \approx 1.5~\mathrm{nm}$, the spin splitting energy suggests that the ripple waves can be utilized to engineer the straintronics devices.   The variation of  strain in GNRs is shown in the inset of  Fig~\ref{fig6}(b) that resembles to the experimentally reported values in Ref.~\onlinecite{bao12} with some deviations to be explained (see the texts for details). We choose $L=3\sqrt{3} ~a N$ ($N=60$) and  $\epsilon_r=2.4$.
}
\end{figure}
\begin{figure}
\includegraphics[width=8.5cm,height=8cm]{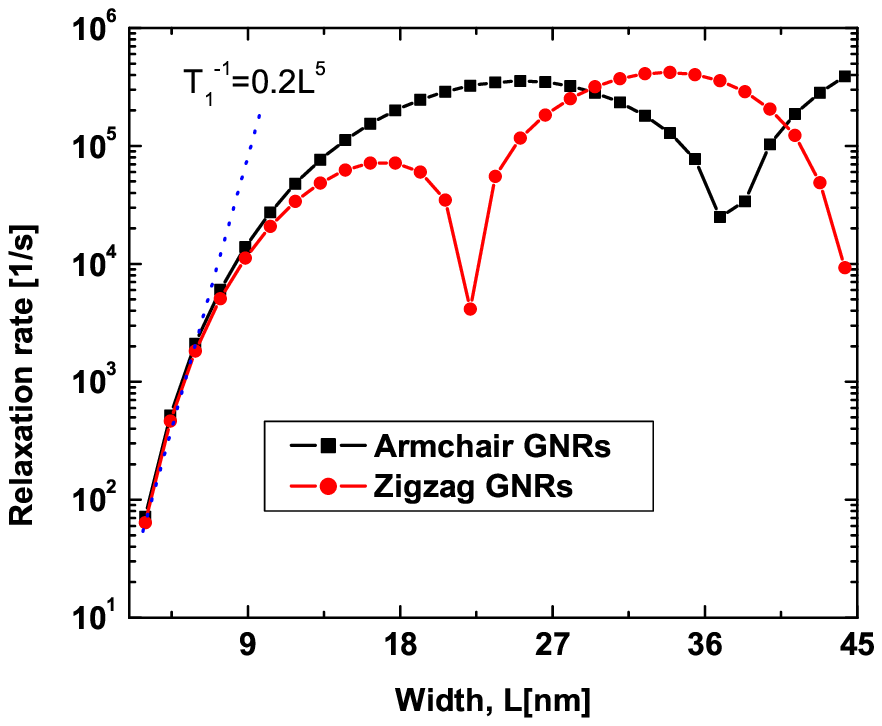}
\caption{\label{fig7}
Electromagnetic field mediated spin relaxation vs width, $L$ at Dirac points in armchair  and zigzag GNRs. The dips at different widths of GNRs is observed due to the fact that the  spin states changes its sign.  Note that the spin relaxation vanishes like $L^5$. We choose $L=3\sqrt{3} ~a N$ ($N=4,6, \cdots$) and $\epsilon_r=2.4$.
}
\end{figure}

\section{Theoretical Model}\label{theoretical-model}
The total elastic energy density  associated with the strain for the two dimensional graphene sheet can be written  as~\cite{landau-lifshitz,carpio08,cadelano12} $2U_s = C_{iklm}\varepsilon_{ik}\varepsilon_{lm}$.
Here $C_{iklm}$ is a tensor of rank four  (the elastic modulus tensor) and  $\varepsilon_{ik}$ (or $\varepsilon_{lm}$) is  the strain tensor.
In the above, the  strain tensor components can be written as
\begin{equation}
\varepsilon_{ik}=\frac{1}{2}\left(\partial_{x_k} u_i+\partial_{x_i} u_k  + \partial_{x_k} h \partial_{x_i} h   \right),\label{varepsilon-ik1}
\end{equation}
where $u_i$ and $h$ are in-plane and out-of-plane displacements, respectively.~\cite{juan13,carpio08}~\cite{bao-lau09,shenoy08,cerda-mahadevan03}
Hence, the strain tensor components for graphene in the 2D  displacement vector $\mathbf{u}(x,y)=(u_x,u_y)$ can be written as
\begin{eqnarray}
\varepsilon_{xx}=\partial_x u_x + \frac{1}{2} \left(\partial_x h\right)^2,\\
\varepsilon_{yy}=\partial_y u_y + \frac{1}{2} \left(\partial_y h\right)^2,\\
\varepsilon_{xy}=\frac{1}{2}\left(\partial_{y} u_x+\partial_{x} u_y\right)+ \frac{1}{2} \left(\partial_x h\right) \left(\partial_y h\right).\label{varepsilon-ik2}
\end{eqnarray}
The stress tensor components $\sigma_{ik}=\partial U_s / \partial \varepsilon_{ik}$ for graphene can be written as~\cite{zhou08}
\begin{eqnarray}
\sigma_{xx}=C_{11}\varepsilon_{xx}+C_{12}\varepsilon_{yy},  \label{sigma-xx}\\
\sigma_{yy}=C_{12}\varepsilon_{xx}+C_{22}\varepsilon_{yy},  \label{sigma-yy}\\
\sigma_{xy}=2C_{66}\varepsilon_{xy}.  \label{sigma-xy}
\end{eqnarray}
In the continuum limit, elastic deformations of graphene sheets are described by the Navier equations $\partial_j \sigma_{ik}=0$. Hence,  the  coupled Navier-type equations of thermoelasticity  for graphene can be written as
\begin{widetext}
\begin{eqnarray}
2\left(C_{11}\partial^2_x+C_{66}\partial^2_y\right)u_x  + 2\left(C_{12}+C_{66}\right) \partial_x \partial_y u_y+
\partial_x\left[C_{11}\left(\partial_x h\right)^2+C_{12}\left(\partial_y h\right)^2\right]+ 2C_{66}\partial_y \left(\partial_x h\right)\left(\partial_y h\right)=0,   \label{coupled-1}\\
2\left(C_{66}\partial^2_x+C_{11}\partial^2_y\right)u_y  + 2\left(C_{12}+C_{66}\right) \partial_x \partial_y u_x +
\partial_y\left[C_{12}\left(\partial_x h\right)^2+ C_{22}\left(\partial_y h\right)^2\right]+ 2 C_{66}\partial_x \left(\partial_x h\right)\left(\partial_y h\right)=0.  \label{coupled-2}
\end{eqnarray}
\end{widetext}
For GNRs  elongated along armchair  direction, only $\varepsilon_{xx}$ is a non-vanishing strain tensor component. Thus, assuming out-of-plane displacement $h=h_x=h_0\sin{\left(2\pi x/\ell\right)}$~\cite{meng13,guinea08a,bao-lau09} along the armchair direction, Eq.~(\ref{coupled-1}) can be simplified as:~\cite{meng13}
\begin{equation}
\partial_x^2 u_x -\frac{4 h_0^2\pi^3}{\ell^3}   \sin\left(\frac{4\pi x}{\ell}\right)=0,   \label{coupled-3}\\
\end{equation}
where $\ell$ is the period and $h_0$ is the height of the ripple waves. For GNRs elongated along the zigzag  direction, only $\varepsilon_{yy}$ is non-vanishing strain tensor component. Thus,  Eq.~(\ref{coupled-2}) can be simplified as:~\cite{meng13}
\begin{equation}
\partial_y^2 u_y -\frac{4 h_0^2\pi^3}{\ell^3}   \sin\left(\frac{4\pi y}{\ell}\right)=0.   \label{coupled-4}\\
\end{equation}
Now we turn to the influence of strain tensor on the electronic properties of GNRs.

In the continuum limit, by expanding the momentum close to the $K$ point in the Brillouin zone, the  Hamiltonian  for $\pi$ electrons at the $\mathbf{\mathrm{K}}$ point reads as:~\cite{maksym13,krueckl12,nato09}
\begin{equation}
H_0= v_F\left(\sigma_x P_x +\sigma_y P_y\right). \label{H0}
\end{equation}
In~(\ref{H0}),  $P=p-\hbar A$ with $p=-i\hbar \partial_x$ being the canonical momentum operator and $\textbf{A}=\beta \left(\varepsilon_{xx}-\varepsilon_{yy},\varepsilon_{xy}\right)/a$ is the vector potential induced by pseudomorphic strain tensor.~\cite{guinea08a,guinea08,juan13} Also,  $a$ is the lattice constant and $\beta=-\partial \ln t/\partial \ln a \approx 2$ with $t$ being the nearest neighbor hoping parameters.

\textbf{Armchair GNRS:} For unstrained graphene nanoribbons with armchair edge,~\cite{sevinifmmode08,zheng07} we assume $\left(H_0\right) \psi=\varepsilon \psi$, where $\psi\left(r\right)=\exp{\left(ik_y y\right)}\left(\phi_A\left(x\right)~\phi_B\left(x\right) \right)^T$.~\cite{nato09} Thus the two coupled equations can be written as
\begin{eqnarray}
-i\hbar v_F \left( \partial_x +k_y\right)\phi_B=\varepsilon \phi_A, \label{hbar-vF-1} \\
-i\hbar v_F \left(\partial_x-k_y\right)\phi_A=\varepsilon \phi_B. \label{hbar-vF-2}
\end{eqnarray}
We apply operator $-i\hbar v_F \left(\partial_x-k_y\right)$ to  Eq.~(\ref{hbar-vF-1}) and by using Eq.~(\ref{hbar-vF-2}), we write second order partial differential equation as:
\begin{equation}
\left(\hbar v_F\right)^2 \left(k_y^2-\partial^2_x\right)\phi_B=\varepsilon^2 \phi_B. \label{hbar-vF-3}
\end{equation}
Its solution is
\begin{equation}
\phi_B=A_1\exp\left(i k_n x\right) + B_1\exp\left(-i k_n x\right), \label{phi}
\end{equation}
where $A_1$ and $B_1$ are arbitrary constant.
One can derive   second order differential equation similar to~(\ref{hbar-vF-3}) at another Dirac point $K'$ and find its solution as
\begin{equation}
\phi'_B=C_1\exp\left(i k_n x\right) + D_1\exp\left(-i k_n x\right), \label{phi}
\end{equation}
where $C_1$ and $D_1$ are another arbitrary constant.
Also, $k_n$ can be found by utilizing boundary conditions  $\psi_A|_{x=0}=\psi_B|_{x=0}=\psi_A|_{x=L}=\psi_B|_{x=L}=0$. To satisfy these boundary conditions, we must chose $A_1=-D_1$ and $B_1=C_1=0$, so we found $k_n$ as:
\begin{equation}
k_n=\frac{n\pi}{L}-\frac{4\pi}{3\sqrt 3 ~a}, \label{kn}
\end{equation}
where $n=0,1,2\cdots $ and   $L$ is the width of the nanoribbons.  We normalized the wavefunctions on each sublattices $A$ and $B$ separately:~\cite{brey06}
\begin{equation}
\int^L_0 dr \left[ |\psi_\mu \left(r\right)|^2  +  |\psi'_\mu \left(r\right)|^2 \right]=1/2.\label{norm}
\end{equation}
Hence, we write the normalized wavefunctions at the Dirac points $K$ and $K'$ as:
\begin{eqnarray}
\phi_A=\frac{\hbar v_F}{2 \varepsilon \sqrt{L}}\left(k_n-ik_y\right)\exp{\left(i k_n x\right)},\label{phi-A-arm}\\
\phi_B=\frac{1}{2\sqrt{L}} \exp{\left(i k_n x\right)},\label{phi-B-arm}\\
\phi'_A=\frac{\hbar v_F}{2 \varepsilon \sqrt{L}}\left(k_n-ik_y\right)\exp{\left(-i k_n x\right)},\label{phi-p-A}\\
\phi'_B=-\frac{1
}{2\sqrt{L}} \exp{\left(-i k_n x\right)}.\label{phi-p-B}
\end{eqnarray}
Finally,  energy eigenvalues of~(\ref{hbar-vF-3}) can be written as
\begin{equation}
\varepsilon_{n\pm}^{a0}=\pm \sqrt {\left(\hbar v_F\right)^2 \left( k_y^2+k_n^2\right) }. \label{varepsilon}
\end{equation}
The wavefunctions and eigenvalues of the unstrained armchair GNRs are shown in Figs.~\ref{fig2} and~\ref{fig3}.

\textbf{Zigzag GNRS:} In a similar way,  GNRs with zigzag edges support two different states such as surface waves (edge states) exist at or near the edge and confined modes. Thus, we write the energy spectrum of the nanoribbons near the edge as:
\begin{equation}
\varepsilon^{z0}_{n\pm}=\pm \sqrt{\left(\hbar v_F\right)^2 \left(k_x^2-z^2\right)},\label{varepsilon-2}
\end{equation}
where $z$ is a real number which  follows the solution of:
\begin{equation}
\exp{\left(-2zL\right)}=\left(k_x-z\right)/\left(k_x+z\right).\label{exp-z}
\end{equation}
For $z=ik_n$, the transcendental Eq.~(\ref{exp-z}) becomes
\begin{equation}
k_x=k_n\cot \left(k_n L\right),\label{k-x}
\end{equation}
and the energy spectrum of GNR for confined modes  is given by
\begin{equation}
\varepsilon^{zc0}_{n\pm}=\pm \sqrt{\left(\hbar v_F\right)^2 \left(k_x^2+k_n^2\right)}.\label{varepsilon-3}
\end{equation}
Also the wavefunctions $ \phi_A(y)$ and $\phi_B(y) $ for GNR with zigzag edge is given by
\begin{eqnarray}
\phi_A=\frac{2iN}{\varepsilon^{zc0}_{n\pm}} \left\{k_x \sin{\left(k_n y\right)} - k_n \cos{\left(k_n y\right)}\right\},\label{phi-A}\\
\phi_B=2iN  \sin{\left(k_n y\right)}.\label{phi-B}
\end{eqnarray}
Since the wavefunctions do not admix the valleys for zigzag GNR, we assume $\langle \phi'_A (y) | \phi'_A (y)\rangle= \langle \phi'_B (y) | \phi'_B (y)\rangle =0$ and  apply the normalization  condition $\langle \phi_A (y) | \phi_A (y)\rangle + \langle \phi_B (y) | \phi_B (y)\rangle =1$ to find the constant $N$ as:
\begin{widetext}
\begin{equation}
|N|^2=\frac{k_n \left(\varepsilon^{zc0}_{n}\right)^2}{\left(k_x^2-k_n^2+ \left(\varepsilon^{zc0}_{n}\right)^2 \right) \left\{ 2 k_n L - \sin\left(2k_n L\right) \right\} +4 k_n \left( k^2_n L - k_x \sin^2{\left( k_nL\right)}  \right) }.
\end{equation}
\end{widetext}
At $k_x=0$,  Eqs.~(\ref{phi-A}) and~(\ref{phi-B}) can be written as
\begin{eqnarray}
\phi_A=\mp \frac{i}{\sqrt {L}} \cos \left[ \frac{\left(2n+1\right)\pi y}{2L}  \right],\label{phi-A-1}\\
\phi_B=\frac{i}{\sqrt {L}} \sin \left[ \frac{\left(2n+1\right)\pi y}{2L}  \right],\label{phi-B-1}
\end{eqnarray}
where $n=0,1,2\cdots$.
The wavefunctions and eigenvalues of the unstrained zigzag GNRs are shown in Figs.~\ref{fig4} and~\ref{fig5}.

\section{Computational Method}\label{computational-details}
The schematic diagram of the two-dimensional  graphene sheet in a computational domain is shown in Fig.~\ref{fig1}.  We have used the multiscale multiphysics simulation framework and solved the Navier's  equations~(\ref{coupled-3}) for armchair GNRs and~(\ref{coupled-4}) for zigzag GNRs via the  finite element method  to investigate the influence of electromechanical effects  on the bandstructures of armchair and zigzag GNRs. By assuming vanishing displacements at the boundary of GNRs, we impose Dirichlet boundary conditions  to find the strain tensor in armchair and zigzag GNRs.

For the bandstructure calculations of GNRs with zigzag edge, we have used Dirichlet boundary conditions for $\phi_A\left(x=0\right)$ and Neumann boundary conditions for $\phi_A\left(x=L\right)$ and viceversa for $\phi_B\left(y\right)$. To find the energy eginvalues of  armchair GNRs, we must admix the two different valleys at Dirac points $K$ and $K'$. Thus, we apply the following conditions at the boundaries of armchair GNRs:
\begin{eqnarray}
\phi_\mu\left(0\right)+\phi'_\mu\left(0\right)=0,\label{phi-mu1}\\
\phi_\mu\left(L\right)+\exp{\left(-2iKL\right)}\phi'_\mu\left(L\right)=0,\label{phi-mu2}
\end{eqnarray}
where $\mu=A,B$.

\section{Results and Discussions}\label{results-discussions}

In Fig.~\ref{fig2}(a), we have plotted the wavefunctions squared vs distance for the armchair GNRs. Here we chose the length of the armchair GNR $L=3\sqrt{3} a N~\left(N=6\right)$. Thus, by considering $n=4N$, we find $k_n=0$ which corresponds to the zero energy eigenvalues and armchair GNRs is considered to show metallic behaviors. In this case, we do not observe any confinement effect in the form of wavefunctions as shown by the dotted line in Fig.~\ref{fig2}(a). For $n=0$ (Fig.~\ref{fig2}(a) (solid line, black)), the wavefunction oscillates  along the armchair GNRs. Fig.~\ref{fig2}(b) illustrates the bandstructure of armchair GNR. As can be seen, the energy difference between electron and hole states at the Dirac point is zero which illustrates the metallic behavior of armchair GNR. In Fig.~\ref{fig3}(a) and (b), we have compared our analytical results  to the numerical values obtained from finite element method by utilizing the boundary conditions~(\ref{phi-mu1}) and~(\ref{phi-mu2}). Here we see that the numerical results are in   excellent agreement to the analytical results.

In Fig.~\ref{fig4}(a), we have plotted the wavefucntions squared vs distance for the lowest energy states of zigzag GNRs for $k_x=0$ and $0.2/nm$.  For the case $k_x=0$, the wavefunctions correspond to the nodeless confined states. This is perfectly described by Eqs.~(\ref{phi-A}),~(\ref{phi-B}),~(\ref{phi-A-1}) and~(\ref{phi-B-1}).
For the case $k_x \neq 0,$ the wavefunctions correspond to the linear combination of the surface states that decay exponentially from the edge. In Fig.~\ref{fig4}(b), we have plotted the bandstructures of zigzag GNRs  and see that the finite width of the GNR breaks the energy spectrum into an infinite set of bands. Here the  solid lines show the  confined modes and dotted lines  show the edge states of the surface waves that has vanishing energy at or  near the edge of the zigzag GNRs. In Fig.~\ref{fig5}(a) and (b), we have compared our analytical results  to the numerical values obtained from finite element method with appropriate boundary conditions: setting the wavefunction  to be zero on the sublattice A on one edge and on the B sublattice on the other. Here again we see that the numerical and analytical results are to be in excellent agreement.

We now turn to the calculations of the transition rate between the two energy levels due to spontaneous emission of photons. We write the total Hamiltonian of graphene under electromagntic field radiation as $H=H_0+H_{int}+H_{so}+ H_A $, where $H_A$ is the additional contribution in the Hamiltonian of graphene due to the electromagentic field radiation of photons  that can be written as
\begin{equation}
H_A=-ev_F \left(\sigma_x A_{x,t} + \sigma_y A_{y,t}\right), \label{HA}
\end{equation}
where $\mathbf{A}\left(\mathbf{r},t\right)$
is the  vector potential   of the electromagnetic field radiation of photons that can be written as
\begin{equation}
\mathbf{A}(\mathbf{r},t)= \sum_{q,\lambda} \sqrt{\frac{\hbar}{2\epsilon_r \omega_{\mathbf{q}}V}} \hat{e}_{q\lambda} b_{q,\lambda} e^{i\left(\mathbf{q\cdot r} -\omega_q t\right)} + H.c.,
\end{equation}
where $\omega_q=c|\mathbf{q}|$,  $b_{q,\lambda}$ annihilate photons with wave vector $\mathbf{q}$, $c$ is the velocity of light, $V$ is the volume and $\epsilon_r$ is the dielectric constant of the graphene nanoribbon. The polarization directions  $\hat{e}_{\mathbf{q}\lambda}$ with $\lambda=1,2$ are chosen as two perpendicular  induced photon modes  in the graphene nanoribbon.  The polarization directions of the induced photon are $\hat{e}_{q1}=\left( \sin\phi,  -\cos\phi, 0 \right)$ and  $\hat{e}_{q2}=\left(\cos\theta \cos\phi, \cos\theta \sin\phi, -\sin\theta \right)$ because we express $\mathbf{q}=q\left(\sin\theta \cos\phi, \sin\theta \sin\phi, \cos\theta \right)$. The above polarization vectors satisfy the relations $\hat{e}_{q1}=\hat{e}_{q2}\times \mathbf{\hat{q}}$, $\hat{e}_{q2}= \mathbf{\hat{q}}\times \hat{e}_{q1}$  and $\mathbf{\hat{q}=\hat{e}_{q1}\times \hat{e}_{q2}}$. Based on the Fermi Golden Rule, the electromagnetic field mediated   transition rate ($i.e.,$ the transition probability per unit time) in the graphene nanoribbon is given by~\cite{merzbacher}
\begin{equation}
\frac{1}{T_1}=\frac{V}{\left(2\pi\right)^2\hbar}\int d^3\mathbf{q}\sum_{\lambda=1,2}\arrowvert M_{q,\lambda}\arrowvert^2\delta\left(\hbar\omega_\mathbf{q}
-\varepsilon_{f}+\varepsilon_{i}\right),
\label{1-T1}
\end{equation}
where  $M_{q,\lambda}=\langle\psi_i|H_A|\psi_f\rangle$. Here  $|\psi_i \rangle$ and $|\psi_f \rangle$ are the initial and final states wavefunctions. By adopting dipole approximation, i.e. the transition is caused only by leading terms from  $A\left(\textbf{r},t\right)$, we write the transition rate as
\begin{equation}
\frac{1}{T_1}=\frac{e^2v_F^2}{4\pi\epsilon_0\epsilon_r\hbar^2c^3}
\left(\varepsilon_f-\varepsilon_i\right).\label{1-T1-1}
\end{equation}
In Fig.~\ref{fig6}, we plot electromagnetic field mediated spin relaxation rate vs height of the ripple waves  in armchair and zigzag GNRs.  By increasing the height of the ripple waves, one can observe an enhancement in the spin splitting energy difference (see inset plot of Fig.~\ref{fig6}(a)) that is similar to Landau levels in semiconductors in the presence of magnetic fields. Such spin splitting due to ripple waves in graphene can be utilized for the potential application in straintronic devices. The spin splitting energy vs height of the ripple waves is shown  in the inset plot of Fig.~\ref{fig6}(a).
Here we find that starting at $h\approx 0.75$, the zero energy eigenvalues disappear, and thus one can find the confinement effect in the form of wavefunctions due to the pshedomorphic  vector potential originating from the electromechanical effects in the armchair GNRs.
The variation of strain in GNRs that is shown in Fig.~\ref{fig6}(b) is similar to Ref.~\onlinecite{bao12} with only a slight deviation due to the fact that we apply vanishing boundary conditions for the displacements at the boundary of the GNRs.
In Fig.~\ref{fig7}, we plot spin relaxation rate vs width of the ripple waves in armchair and zigzag GNRs. Here we see that, in both armchair and zigzag GNRs, the spin relaxation rate varies like $L^5$. It can be seen that one can engineer the variation of the strain in such as way that the sign change in the spin flip behavior in the manipulation of spin relaxation can induce the dips  at the widths $L=22\mathrm{nm}$ for zigzag GNRs and at $L=36\mathrm{nm}$ for armchair GNRs. Similar  spin flip behaviors i.e. sign change in the effective g-factor (see Refs.~\onlinecite{prabhakar09,prabhakar11,prabhakar12, prabhakar13}) can be observed with the application of gate controlled electric fields  in III-V semiconductor quantum dots.

\section{Conclusion}\label{conclusion}

Based on analytical and finite element numerical results, from Figs.~\ref{fig2} to~\ref{fig5}, we have analyzed the bandstructures of armchair and zigzag graphene nanoribbons. In particular, we have shown that the armchair GNRs with zero energy eigenvalues exhibit  metallic behaviors, see Figs.~\ref{fig2} and~\ref{fig3}. The metallic behavior turns into the semiconducting approximately at $h_0=2 \mathrm{nm}$, Fig.~\ref{fig6}(a), due to ripple waves originating from electromechanical effects for straintronic applications. In Fig.~\ref{fig6}, we have shown that the ripple waves originating from electromechanical effects can be treated as a phedomorphic vector potential in the bandstructure calculation of GNRs that provides the spin splitting energy similar to Zemann type Landau energy levels in semiconductors. Such spin splitting energy allowed us to investigate the spin relaxation behaviors in armchair and zigzag GNRs. Estimated numerical values of the spin relaxation time is in agreement to the experimental results. Finally, in Fig.~\ref{fig7}, we have shown that the spin relaxation rate varies like $L^5$ in smaller width  GNRs. In this figure, we have also shown that it is possible to tune the spin currents ON and OFF in the manipulation of spin flip behaviors  due to ripple waves at different widths of GNRs  that might have the potential applications in straintronic devices. We have shown that such finding is  similar to the recently reported sign change behaviors in the effective Land$\mathrm{\acute{e}}$ g-factor in III-V semiconductor quantum dots.

\begin{acknowledgments}
This work has been supported by NSERC and CRC programs (Canada). The authors  acknowledge the Shared Hierarchical Academic Research Computing Network (SHARCNET) community  and Dr. P.J.  Douglas Roberts for his assistance and technical support.
\end{acknowledgments}

%

\end{document}